# The effect of LPSO phase on the high-temperature oxidation of a stainless Mg-Y-Al alloy

Zhipeng Wang[1], Zhao Shen[1*], Yang Liu[1], Yahuan Zhao[1], Qingchun Zhu[1], Yiwen Chen[1], Jingya Wang[1], Yangxin Li[1*], Sergio Lozano-Perez[2], Xiaoqin Zeng[1*]

[1] National Engineering Research Center of Light Alloy Net Forming and State Key Laboratory of Metal Matrix Composites, School of Materials Science and Engineering, Shanghai Jiao Tong University, Shanghai 200240, China

[2] Department of Materials Science, University of Oxford, Parks Road, Oxford OX1 3PH, UK

Corresponding authors: Z. Shen shenzhao081@sjtu.edu.cn; Y. Li astatium@sjtu.edu.cn; X. Zeng xqzeng@sjtu.edu.cn;

**Abstract**: In this study, we investigated the oxidation of the Mg-11Y-1Al alloy at 500°C in an Ar-20%$O_2$ environment. Multiscale analysis showed the network-like long-period stacking ordered (LPSO) phase transformed into needle-like LPSO and polygonal $Mg_{24}Y_5$ phases, leading to the formation of a high-dense network of needle-like oxides at the oxidation front. These oxides grew laterally along the oxide/matrix interfaces, forming a thicker, continuous scale that effectively blocked elemental diffusion. Hence, the preferential oxidation along the needle-like LPSO is believed to accelerate the formation of a thicker and continuous oxide scale, further improving the oxidation resistance of the Mg-11Y-1Al alloy.

**Keywords:** Mg alloy; LPSO phase; Oxidation; TEM; TKD.

Mg alloys, known for being the lightest structural metallic materials, hold significant potential in energy-saving applications due to their role in reducing overall load [1-5]. These alloys are increasingly sought after in industrial applications, prized for their unique combination of low density, good castability, excellent recyclability, and high biodegradability, among other qualities [2, 3, 6, 7]. Despite these advantages, a notable limitation of Mg alloys is their generally insufficient mechanical strength. To overcome this challenge, a variety of Mg alloys have been developed, particularly those incorporating the LPSO phase. Alloys with LPSO have been shown to exhibit remarkable improvements in mechanical properties [8-10]. The LPSO phase, as a significant and specialized precipitate in Mg alloys, contributes to enhanced strength, stiffness, and creep resistance [11-14]. In recent advancements, our research group has successfully developed a new stainless Mg-11Y-1Al (wt.%) alloy that includes the LPSO phase [15]. This innovation has led to a dual benefit: not only does the LPSO phase increase the mechanical strength of the alloy, but it also modulates the corrosion behavior. We found that the inclusion of the LPSO phase shifts the corrosion pattern from pitting to uniform corrosion, effectively reducing the overall corrosion rate of the alloy [15].

Mg alloys are generally perceived to have poor high-temperature oxidation resistance due to their strong affinity for oxygen and the absence of a protective oxide scale [16-20]. This susceptibility becomes particularly problematic during high-temperature manufacturing processes such as extrusion, welding, forging, and heat treatment, where Mg alloys are at risk of severe oxidation or even combustion [16, 17, 21-23], thus limiting their commercial viability and potential applications. Therefore, understanding the oxidation behaviors of stainless Mg-11Y-1Al alloy at elevated temperatures is crucial to expanding its commercial use. Previous studies indicate that precipitates rich in rare-earth elements can significantly impair the high-temperature oxidation performance of Mg alloys [23-25]. However, the specific impact of the LPSO phase on the oxidation behaviors of Mg-11Y-1Al alloy at high temperatures remains unclear. In this study, we investigate the oxidation kinetics and oxide scale evolution of Mg-11Y-1Al alloy, containing a substantial amount of LPSO precipitates, using advanced characterization techniques.

Commercially pure Mg, Mg-25Y master alloy, and pure Al were melted in an electric resistance furnace to synthesize a Mg-11Y-1Al alloy (wt.%) [15, 26]. The actual chemical composition tested by Inductive Coupled Plasma Emission Spectrometer (ICP) was Mg-10.83Y-1.09Al (wt.%). For high-temperature testing, samples measuring 15 × 15 × 3 mm³ were meticulously polished with a 0.05 μm diamond suspension. The testing was conducted in an Ar-20%$O_2$ atmosphere at 500°C. Weight measurements of the samples were taken at specific intervals—1, 20, 60, 100, and 150 h—during the oxidation testing. Nine samples were utilized in the high-temperature oxidation test. At designated sampling intervals, every sample was extracted for weight assessment. To ascertain the mean weight of each sample, it was subjected to five separate measurements. Post-testing, the surface and cross-sectional morphologies of the specimens were thoroughly examined using a scanning electron microscope (SEM) equipped with both an energy-dispersive X-ray (EDX) detector and an electron backscattered diffraction (EBSD) detector. For more detailed analysis, site-specific thin foils were prepared using a focused ion beam (FIB) and then analyzed using a transmission electron microscope (TEM) and transmission Kikuchi diffraction (TKD).

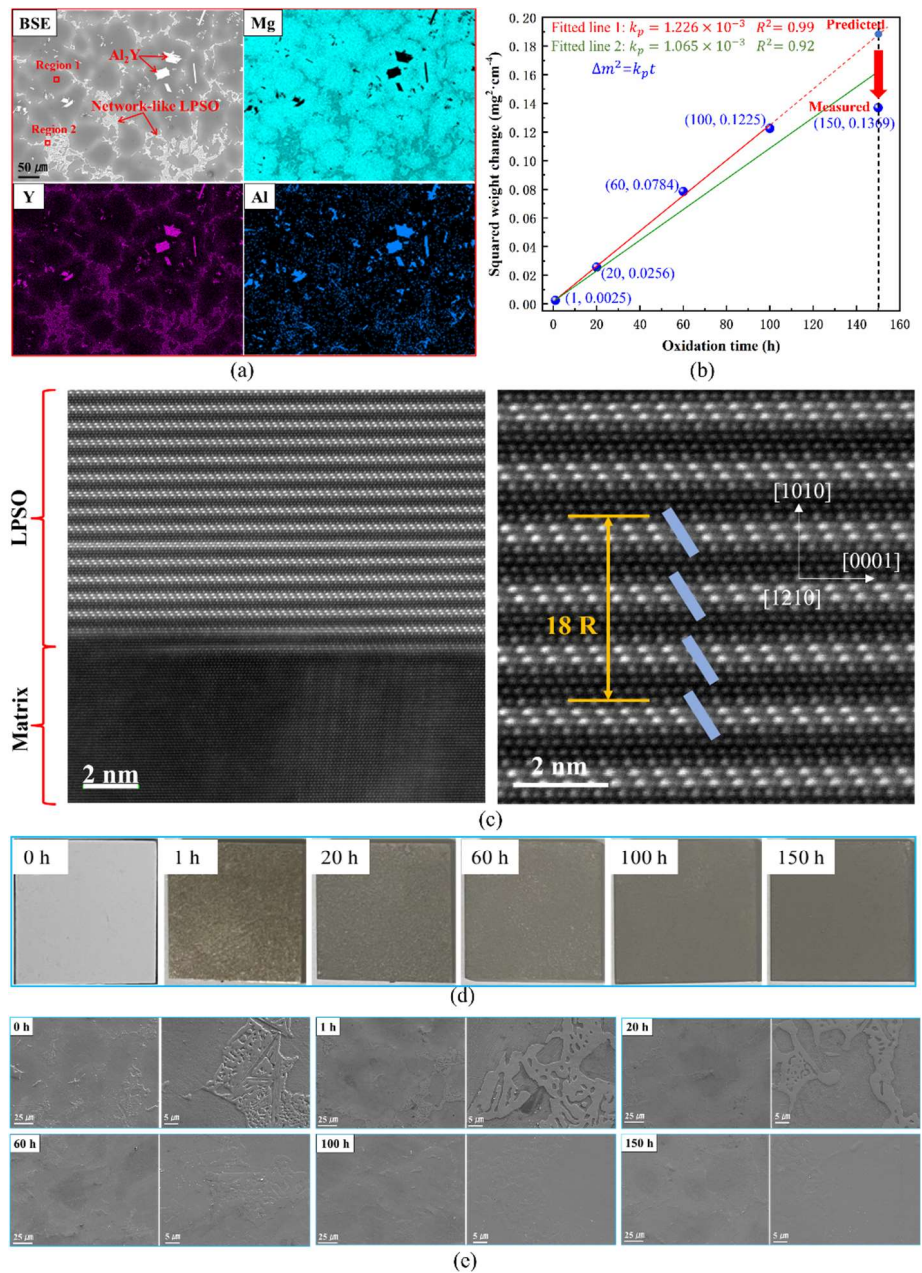

Fig. 1. (a) BSE and EDX images of Mg-11Y-1Al alloy prior to oxidation testing; (b) weight gain curve; (c) atomic-resolution HAADF image of LPSO and matrix; (d) macrographs and (e) SEM-SE images of the Mg-11Y-1Al alloy before and after the oxidation testing at 500 °C for different times.

Fig. 1a presents the backscattered electron (BSE) images and corresponding energy-dispersive X-ray (EDX) elemental distribution images of the Mg-11Y-1Al alloy before undergoing oxidation testing. These images confirm the presence of $Al_2Y$ and LPSO secondary phases within the alloy substrate [15]. In the BSE image, the brighter polygonal particles are identified as the $Al_2Y$ phase, distinguished by their higher Y content [8, 26, 27]. The less bright, network-like precipitates are the LPSO phase, comprising Mg, Y, and Al, as verified by the elemental distribution maps. There are also a very few $Mg_{24}Y_5$ phase formed in Mg-11Y-1Al. The chemical compositions (wt.%) in the typical matrix and LPSO are measured by SEM-EDX, which are Mg-2.1%Y-0.2%Al (region 1 in Fig. 1a) and Mg-10.6%Y-7.3%Al (region 2 in Fig. 1a), respectively. The atomic-resolution high-angle annular dark-field (HAADF) imaging was conducted to examine the phase structure of the LPSO. Fig. 1c shows that the examined LPSO has a typical 18R structure, which is consistent with the results in literature [8, 9]. The volume fraction of the $Al_2Y$ phase in the alloy is considerably smaller (~0.3%) compared to the network-like LPSO phase (~21%). Consequently, the influence of the $Al_2Y$ phase on the alloy oxidation is considered negligible and is not the central focus of this study. Fig. 1b illustrates the weight gain spots of the Mg-11Y-1Al alloy after undergoing oxidation at 500 °C for various durations. The fitted line-1 indicates that this alloy had experienced a typical parabolic oxidation up to 100 h, which can be properly fitted by a parabolic equation [1], as shown below:

$$\Delta m = (k_p \cdot t)^{0.5} \tag{1}$$

$\Delta m$ is the weight gain of the alloy per unit area (mg/cm$^2$); $k_p$ is the oxidation rate constant (mg$^2$·cm$^{-4}$·h$^{-1}$); $t$ represents the oxidation time (h). The calculated rate constant $k_p$-1 was $1.226 \times 10^{-3}$. The fitted line-2 indicates that the oxidation did not strictly follow a parabolic law during the entire oxidation duration, which is mainly due to the weight gain at 150 h is much lower than that required for a parabolic kinetics. Fig. 1d reveals the surface conditions of the original Mg-11Y-1Al alloy specimen before and after the oxidation testing. Initially, the surfaces were extremely smooth due to fine polishing. However, noticeable coarsening of the surfaces occurred after just 1 h of oxidation. Interestingly, as the oxidation time increased, the surfaces tended to become smoother again. This coarsening indicates that the early stages of oxidation were not uniform, leading to uneven surface textures. As oxidation progressed, more uniform oxide scales formed, contributing to the eventual smoothening of the surfaces. The SEM-SE images provide further insight (Fig. 1e). Initially, the positions of the LPSO phase within the alloy were clearly visible after 1 h of oxidation. However, with prolonged oxidation, it became increasingly difficult to discern the LPSO phase locations, and after 150 h, they were almost indiscernible. These observations explain why the surfaces tended to be smoother over time. This transition is indicative of the evolving nature of the oxidation process under extended exposure of this Mg alloy.

Fig. 2a displays the SEM-BSE images of the Mg-11Y-1Al alloy before and after 20 h of oxidation testing. The network-like LPSO phase remains observable after oxidation. The EDX images around the LPSO phase, as seen in Fig. 2b, indicate that it is enriched in Y and Al, consistent with observations made in Fig. 1a. However, closer examination reveals that part of the network-like LPSO phase decomposed under the high-temperature conditions, leading to the formation of needle-like LPSO phase and polygonal $Mg_{24}Y_5$ phase (Fig. 2c). This observation aligns with findings by Xu et al. [28] in their study of a Mg-Y-Zn alloy oxidized at elevated temperatures. Moreover, a recent study by Chen et al. [9] on LPSO phase equilibria in Mg-Y-Al alloys at 450-550 °C also noted the decomposition of the network-like LPSO phase into needle-like LPSO and polygonal $Mg_{24}Y_5$ phases. It was reported that the network-like LPSO phase completely decomposed after annealing for 10 days.

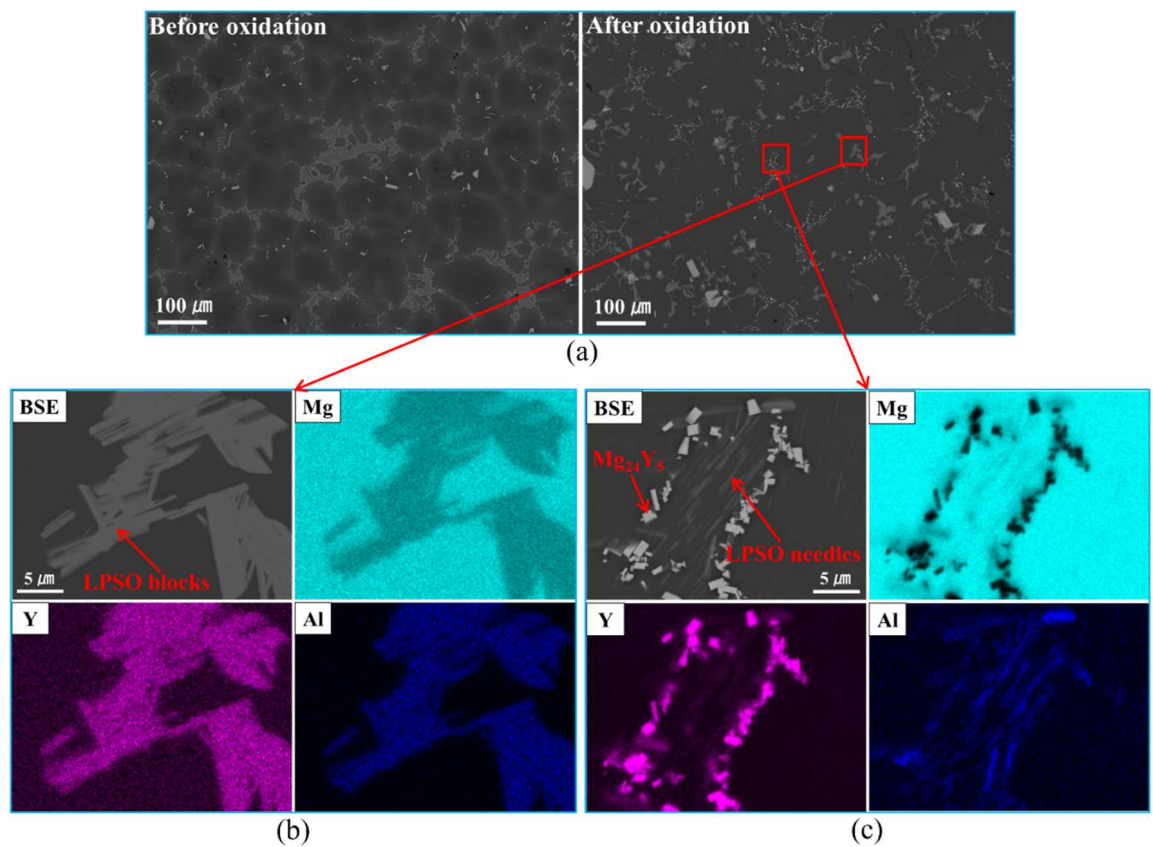

Fig. 2. (a) BSE images showing the matrix morphologies of Mg-11Y-1Al alloy before and after 20 h of oxidation; BSE and the corresponding EDX images of the network-like LPSO (b) before and (c) after decomposition.

Fig. 3 showcases the cross-sectional morphologies of the Mg-11Y-1Al alloy after varying durations of oxidation. Notably, a thin and continuous oxide scale formed on the specimen surface after just 1 h of exposure (Fig. 3a). The oxide scale thickness was about 200 nm on the α-Mg matrix, while it was approximately 400 nm on areas initially containing the network-like LPSO phase. This discrepancy in oxidation rates between the α-Mg matrix and the network-like LPSO phase could account for the initially coarse surface morphology observed. Further analysis revealed that the near-surface network-like LPSO phase began decomposing as the exposure time increased. This decomposition led to the formation of needle-like LPSO phase and polygonal $Mg_{24}Y_5$ phase (Figs. 3a-d). Given that the polygonal $Mg_{24}Y_5$ phase is not thermally stable [25], it gradually decomposed and dissolved into the α-Mg matrix with extended exposure time. Moreover, the thicknesses of the oxide scales on both the α-Mg matrix and the network-like LPSO phase progressively increased with longer oxidation times. Intriguingly, the cross-sectional morphologies of the oxide scales varied between these two regions. The oxide scale on the α-Mg matrix appeared generally uniform, whereas a high density of needle-like oxides was observed at the oxidation front of the network-like LPSO phase.

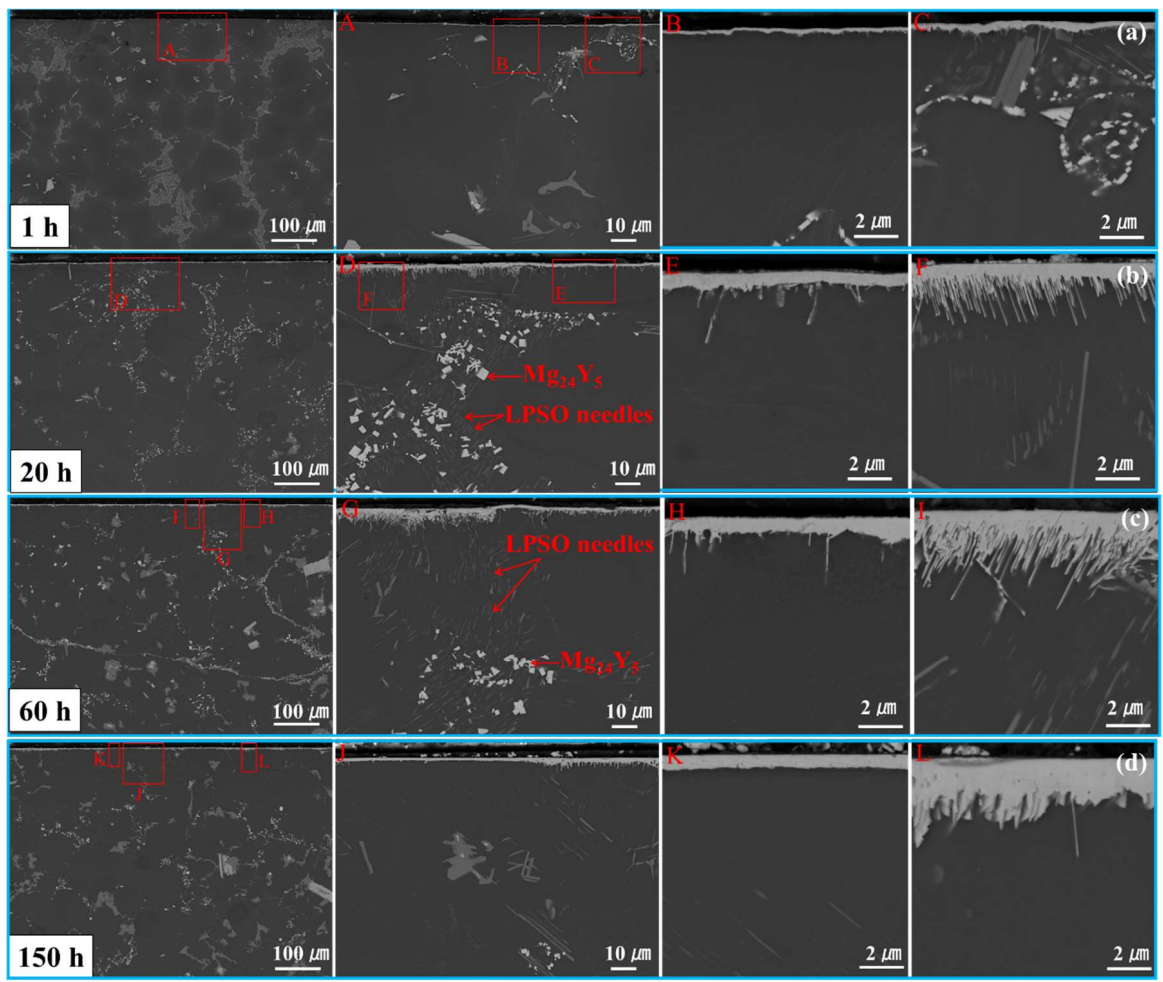

Fig. 3. SEM-BSE images showing the cross-sectional morphologies of the Mg-11Y-1Al alloy after oxidation for (a) 1 h; (b) 20 h; (c) 60 h; (d) 150 h.

To elucidate the microstructure and microchemistry of the oxide scales on the α-Mg matrix and the network-like LPSO phase, two targeted TEM foils were crafted from a specimen after 20 h of exposure. TEM foil-1, extracted from the α-Mg matrix, was analyzed using high-angle annular dark-field (HAADF) imaging and Energy-Dispersive X-ray (EDX) mapping, as depicted in Fig. 4a. These analyses revealed that the oxide scale on the α-Mg matrix was rich in Y and Al, but showed a depletion in Mg. The distribution of Y and Al appeared uniform. EDX line profiles suggested that this oxide scale was predominantly composed of $Y_2O_3$, as inferred from Fig. 4b. The chemical composition analysis at designated Area 1 and Area 2 indicated a significantly higher Al content in the oxide scale compared to the underlying α-Mg matrix (~1.1 vs. ~0.1 at.%). To further decipher the phase structure of the oxide scale, selective area electron diffraction (SAED) was performed. The SAED pattern, shown in Fig. 4c, confirmed the presence of $Y_2O_3$. Notably, despite the presence of ~1.1 at.% Al in the oxide scale, $Al_2O_3$ was not detected in the SAED analysis. Additionally, transmission Kikuchi diffraction (TKD) analysis was conducted on TEM foil-1 (also illustrated in Fig. 4c), revealing that the oxide scale consisted of columnar $Y_2O_3$ grains.

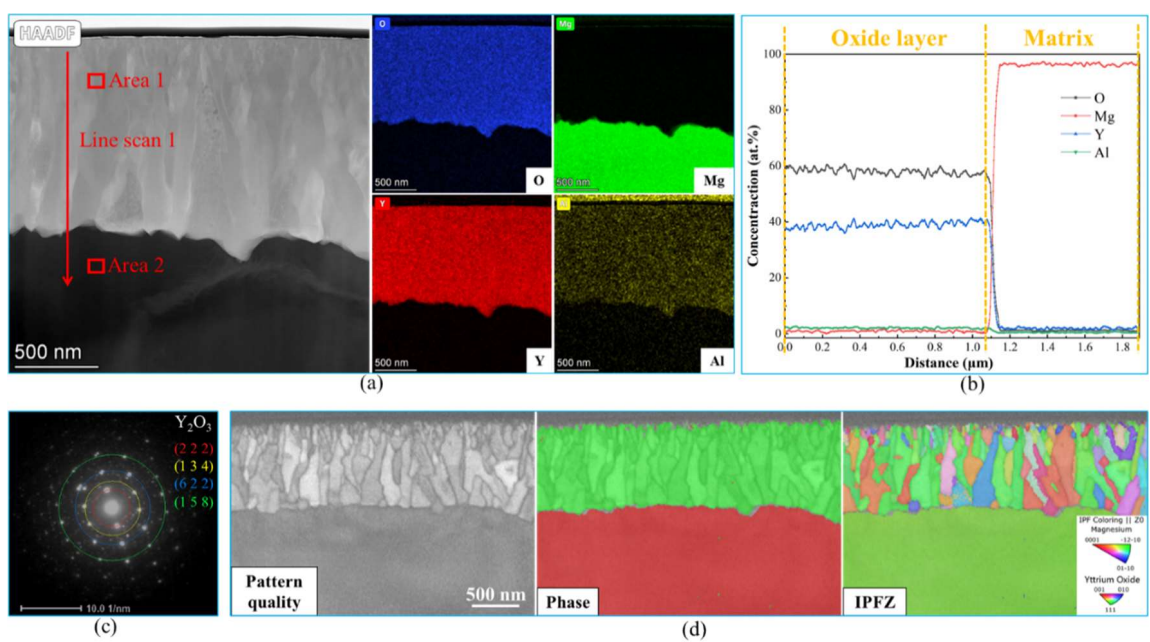

Fig. 4. (a) STEM-HAADF and EDX images of the oxide scale on matrix after oxidation; (b) EDX line-scan 1 exhibiting the distribution of elements through the oxide scale; (c) SAED patterns of oxide scale; (d) TKD results of the oxide (Mg matrix in red, $Y_2O_3$ in green, step size = 5 nm)

TEM foil-2 was meticulously prepared from the oxidation front of the network-like LPSO phase following 20 h of oxidation. This area was characterized by a high density of needle-like oxides. The HAADF imaging and EDX mapping, as illustrated in Fig. 5a, indicated that these needle-like oxides were also enriched in Y and Al, but depleted in Mg. The EDX line profiles, showcased in Fig. 5b, along with the high-resolution Transmission Electron Microscopy (HRTEM) image, displayed in Fig. 5c, confirmed that the needle-like oxide primarily consisted of $Y_2O_3$. The analysis of the chemical composition in specified areas (Area 3 and Area 4) revealed Al contents of ~1.3 and ~0.2 at.%, respectively. These measurements were found to be strikingly similar to those observed in TEM foil-1. TKD analysis further substantiated that the needle-like oxide was composed of $Y_2O_3$. A notable observation was that these needle-like oxides were identified as bicrystals in two dimensions (2D), as indicated by the gathered data.

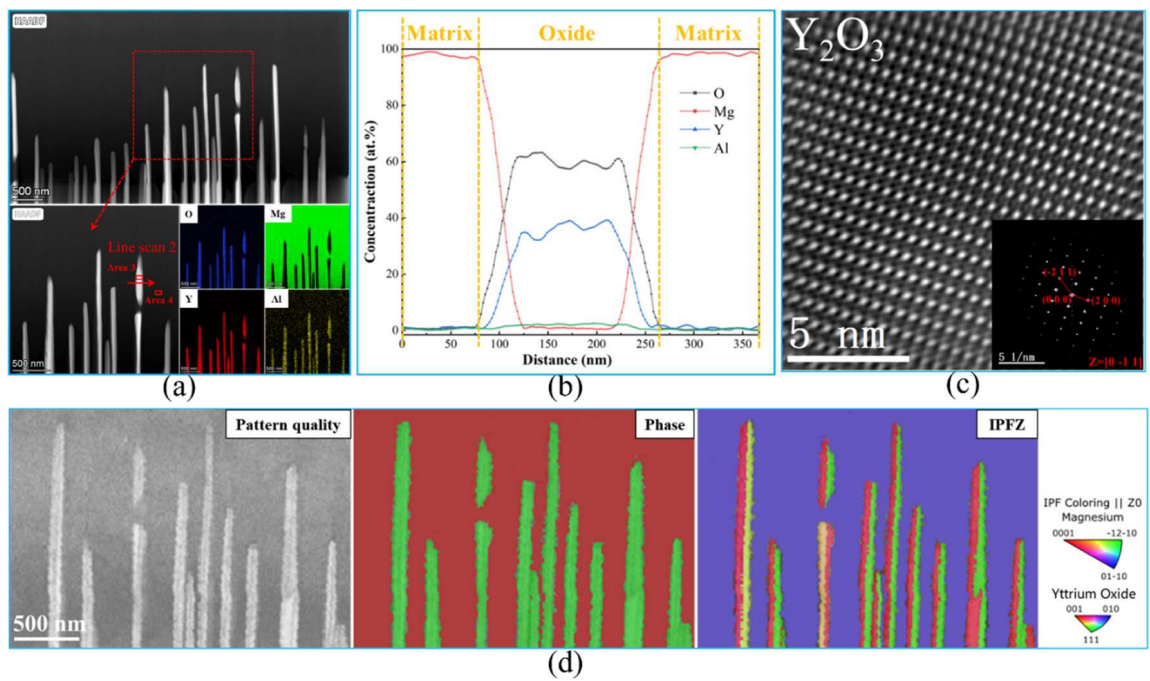

Fig. 5. (a) STEM-HAADF and EDX images of the needle-oxides after oxidation; (b) EDX line-scan 2 exhibiting the distribution of elements through needle-oxide; (c) HRTEM image of needle-oxide; (d) TKD results of needle-like oxides (Mg matrix in red, $Y_2O_3$ in green, step size = 5 nm)

The findings outlined above reveal that the oxide scales developed on the Mg-11Y-1Al alloy were initially continuous and compact (as seen in Fig. 3a). Both on the α-Mg matrix and the network-like LPSO phase, the oxide scales primarily consisted of $Y_2O_3$. $Y_2O_3$ has been extensively recognized as an effective diffusion barrier, proficient in impeding both inward oxygen and outward metallic ion diffusion [25, 29-31]. Interestingly, the oxide scale on the network-like LPSO phase exhibited a greater thickness compared to that on the α-Mg matrix. This indicates a higher oxidation rate in the network-like LPSO phase in the initial stages of the process. As oxidation progressed, the oxide scales on the α-Mg matrix tended to thicken gradually. Concurrently, needle-like oxides began to appear at the oxidation front of the network-like LPSO phase. The emergence of these needle-like oxides is indicative of accelerated oxidation occurring along specific directions. The specific directions are the interfaces between these needle-like oxides and the α-Mg matrix, these interfaces would act as rapid pathways for further inward oxygen diffusion.

In one of our recent studies [23], we discovered that the Mg-Nd secondary phase negatively impacts the oxidation resistance of a ZM6 alloy. Drawing from both literature [9, 26] and our current findings (as shown in Figs. 2 and 3), it was observed that the network-like LPSO phase was unstable at 500 °C. Over time, it progressively transformed into needle-like LPSO and polygonal $Mg_{24}Y_5$ phases. The polygonal $Mg_{24}Y_5$ phase, however, was not thermally stable at this temperature and subsequently decomposed, integrating into the α-Mg matrix [25]. In contrast, the needle-like LPSO phase was found to be thermostable at 500 °C [9]. As the network-like LPSO phase decomposed, further oxidation exposed the newly formed needle-like LPSO phase to inward oxygen diffusion. The standard Gibbs free energies for the oxidation reactions of Mg and Y at 500 °C are -1038987 and -1120536 [J/mol of $O_2$], respectively [32]. This indicates that Y is more reactive in oxidation processes than Mg at this temperature [33]. Both the network-like LPSO phase and the needle-like LPSO phase, as seen in Figs. 2b and c, were enriched in Y compared to the α-Mg matrix. This higher Y content in the LPSO phase likely contributes to their faster oxidation rates compared to the α-Mg matrix, as evidenced in Fig. 3a. Furthermore, the emergence of high-density needle-like oxides at the oxidation front of the network-like LPSO phase after prolonged exposure can be attributed to accelerated in-situ oxidation along the needle-like LPSO phase.

The investigation revealed that all the needle-like oxides were bicrystals in 2D, as shown in Fig. 5d. The formation of these bicrystals is thought to result from accelerated inward oxygen diffusion along the interfaces between the needle-like LPSO and α-Mg matrix and oxide nucleation from these interfaces to the center of needle-like LPSO. Previous studies have indicated that the LPSO phase in the Mg-11Y-1Al alloy possesses a monoclinic structure [8, 9]. This suggests the possibility of multiple LPSO/α-Mg matrix interfaces in a three-dimensional (3D) context. Therefore, it can be inferred that the needle-like $Y_2O_3$ might comprise more than two grains when viewed in 3D. Once the needle-like $Y_2O_3$ formed, the interfaces between these needle-like oxides and the α-Mg matrix would act as rapid pathways for diffusion. This would facilitate further inward oxygen diffusion along these interfaces, reacting with Y in the surrounding α-Mg matrix and leading to the lateral expansion of these needle-like oxides. After 150 h of oxidation, the initially discrete needle-like $Y_2O_3$ oxides became interconnected, forming a thicker and continuous $Y_2O_3$ oxide scale, as depicted in Fig. 3d. It is postulated that this thicker oxide scale, developed on the network-like LPSO phase after prolonged oxidation, could be more effective in obstructing elemental diffusion. This hypothesis could be supported by the observation of the deviation from parabolic kinetics at 150 h (Fig. 1b). The decrease of the weight gain rate during this period might be because the needle-like $Y_2O_3$ oxides became interconnected and effectively increase the protectiveness of the oxide scale.

Al was deliberately added to Mg alloys in the hopes of forming a protective $Al_2O_3$ oxide scale at elevated temperatures. However, contrary to expectations, this led to the formation of a low-melting-point (437°C) β-$Mg_{17}Al_{12}$ phase, which significantly exacerbated oxidation and the evaporation of Mg at elevated temperatures [25, 27]. In the current study, the Al-rich precipitates were found to be the high-melting-temperature $Al_2Y$ phase, not the low-melting-point β-$Mg_{17}Al_{12}$ phase [25, 27]. This discovery meant that the previously observed adverse effects were not present in this alloy. Although the oxides formed on the Mg-11Y-1Al alloy were predominantly $Y_2O_3$, ~1 at.% of Al was detected within these oxides (as shown in Figs. 4a and 5a). The standard Gibbs free energy for the oxidation reaction of Al at 500 °C is reported as -955918 [J/mol of $O_2$] [33], suggesting that Mg is more reactive in oxidation than Al. Given that Mg was not oxidized in this study, it is believed that the Al present in the $Y_2O_3$ oxides existed in a solid-solution state rather than as $Al_2O_3$. It is proposed that Al might partially incorporate into the $Y_2O_3$ oxide scale, forming a (Y,Al)O solid solution. This incorporation can induce lattice distortions and hinder dislocation movement, resulting in significant solid solution strengthening. Therefore, the strengthened (Y,Al)O oxide scale is likely to exhibit increased resistance to cracking. This enhanced resistance could delay the onset of cracking and debonding of the surface oxide scale during high-temperature testing. Similar phenomena, such as the strengthened (Mg,Be)O oxide scale, have been observed and experimentally verified in other studies [34-36].

In summary, the Mg-11Y-1Al alloy, when exposed to an Ar-20%$O_2$ environment at 500 °C, demonstrated effective protective oxidation, characterized by a parabolic during the period of 0 to 100 h. This oxidation resistance is primarily due to the development of a continuous and dense $Y_2O_3$ oxide scale. Despite the transformation of the network-like LPSO phase into needle-like LPSO and polygonal $Mg_{24}Y_5$ phases, the surface oxide scale of the Mg-11Y-1Al alloy remained intact. The expedited in-situ oxidation of the needle-like LPSO phase contributed to the formation of a thicker and more continuous $Y_2O_3$ oxide scale after 150 h of exposure. This scale is considered to be a more efficient diffusion barrier compared to that formed on the α-Mg matrix, which explains the decreased weight gain rate during the period of 100 to 150 h.

## Acknowledgements

The authors would like to thank the financial support from the National Key Research and Development Program of China (No. 2022YFB3708400), the National Science and Technology Major Project (J2019-VIII-0003-0165), and the Space Utilization System of China Manned Space Engineering (No. KJZ-YY-WCL04).